\documentclass[aps,prl,twocolumn,superscriptaddress]{revtex4}
\usepackage{amsmath,epsfig,mathrsfs,graphicx,amssymb}

\def\be#1\ee{\begin{equation}#1\end{equation}}

\newcommand{\ba}{\begin{eqnarray} }
\newcommand{\ea}{\end{eqnarray} }
\begin{document}

\title{Quasiprobabilistic Interpretation of Weak measurements in
  Mesoscopic Junctions}

\author{Adam Bednorz}
\email{Adam.Bednorz@fuw.edu.pl}
\affiliation{Fachbereich Physik, Universit{\"a}t Konstanz, D-78457 Konstanz, Germany}
\affiliation{Institute of Theoretical Physics, University of Warsaw, Ho\.za 69, PL-00681 Warsaw, Poland}
\author{Wolfgang Belzig}
\affiliation{Fachbereich Physik, Universit{\"a}t Konstanz, D-78457 Konstanz, Germany}
\date{\today}

\begin{abstract}
  The impossibility of measuring noncommuting quantum mechanical
  observables is one of the most fascinating consequences of the
  quantum mechanical postulates. Hence, to date the investigation of
  quantum measurement and projection is a
  fundamentally interesting topic. We propose to test the concept of
  weak measurement of noncommuting observables in mesoscopic transport experiments,
  using a quasiprobabilistic description. We derive an
  inequality for current correlators, which is satisfied by every classical probability
  but violated by high-frequency fourth-order cumulants in the quantum
  regime for experimentally feasible parameters.
  

\end{abstract}

\maketitle

Every measurement in quantum mechanics is in principle described by
the projection postulate \cite{neumann}.  However, in practice perfect projective
detectors often do not exist and the measurement encounters a finite
error.  This can be resolved by replacing the projection
by Kraus operators defining a positive operator-valued measure
(POVM)\cite{kraus,povm}. The Kraus operator can be continuously
changed from projection -- strong measurement (exact) -- to almost
identity operator -- weak measurement (with huge random
error). Effectively, a POVM means that we take detector's degrees of
freedom as part of the considered Hilbert space and make a projective
measurement on the detector. Obviously, in that case the
detector-system coupling defines the strength of the measurement of
the system. The equivalence of a POVM and the projective measurement
follows from Naimark theorem \cite{naim}. The actual modeling of
a detection scheme by POVM is a long-standing problem
\cite{brag,jordan,les08}.

The interpretation of the results of a weak measurement can lead to
paradoxes.  For instance, if a weak measurement of $\hat{A}$ performed on the state
$\hat{\rho}$ is followed by a projection $\hat{B}$ then the \emph{weak
  value} can be defined 
$ {}_B\langle A\rangle_{\rho}=
\mathrm{Tr}(\hat{B}\hat{A}\hat{\rho})/\mathrm{Tr}(\hat{B}\hat{\rho})
$ \cite{aav}.
The unusual feature of the weak value is that it can exceed the
spectrum of $\hat{A}$, which obviously contradicts our classical
intuition.  The
strange properties of weak measurements have been confirmed
experimentally in quantum optics \cite{wvopt}, while experiments in
solid state physics are proposed \cite{wvcond}. The interpretation of recent experiments on current fluctuations in
mesoscopic junctions in the quantum regime \cite{glatt-gab}
is impossible in terms of a usual probability \cite{ours}.  
Instead, we proposed to consider a weak
current measurement, which implies a large background noise, but
avoids the paradox of a certain average square to become negative.
The necessity of a weak measurement lead to corrections in the
observed finite-frequency noise, however, a direct experimental proof
of this scheme was not feasible.

In this Letter, we first construct a general formula for a quantum
quasiprobability, which does not depend on the details of the
measurement apparatus and confirms the previously used formulas
\cite{ours}. Second, we propose a scheme to test experimentally its
negativity in frequency domain.  To this end, we will derive a classical inequality for
high-frequency current correlators of the form
\begin{equation}
  \label{eq:1}
  C_{\omega\omega'}^2\leq
  \left(C_{\omega\omega}+2\pi S_\omega^2/\delta\omega\right)
  \left(C_{\omega'\omega'}+2\pi S_{\omega'}^2/\delta\omega'\right)\,.
\end{equation}
Here $C_{\omega\omega'}=\langle\langle I(\omega)
I(-\omega)I(\omega')I(-\omega')\rangle\rangle/t_0$ is a
fourth-order correlator and $S_\omega=\langle\langle
I(\omega)I(-\omega)\rangle\rangle/t_0$ the frequency-dependent current
noise, where $t_0$ is the total (long) measurement time and
$\delta\omega$ is the bandwidth of the detector ($S$ and $C$ are independent of $t_0$). Inequality
(\ref{eq:1}) is satisfied by every classical stochastic process, but
can be violated by high-frequency correlators in the quantum regime of
a mesoscopic junction for experimentally accessible parameters. We
believe our proposed violation of the classical
inequality (\ref{eq:1}) can be realized with the existing techniques 
\cite{glatt-gab}. This violation will be a proof of negative
values of the quasiprobability.  Although it is not necessary to
explain the strange features of weak values \cite{aav}, it offers an
alternative test of nonclassicality similar to the Wigner function
\cite{wigner}.  Moreover, the quasiprobabilistic interpretation can be
easily generalized to an arbitrary sequence of measurements.  This
interpretation facilitates the transfer to mesoscopic junctions and we
present an example, how the negativity of the quasiprobability can be
proven in a tunnel contact and discuss the experimental
feasibility.

We will construct the quasiprobability by a deconvolution from a
suitable POVM. The real parts of weak values can then be expressed as
averages with respect to the quasiprobability.  Let us begin with the
basic properties of a POVM. The Kraus operators $\hat{K}(A)$ for an
observable described by $\hat{A}$ with continuous outcome $A$ need
only to satisfy $\int dA\hat{K}^\dag(A)\hat{K}(A)=\hat 1$. The act of
measurement on the state defined by the density matrix $\hat{\rho}$
results in the new state $\hat{\rho}(A) =
\hat{K}(A)\hat{\rho}\hat{K}^\dag(A)$. The new state yields a
normalized and positive definite probability density $\rho(A)=\mathrm{Tr}\,\hat{\rho}(A)$.  The
procedure can be repeated recursively for an arbitrary sequence of (not
necessarily commuting) operators $\hat{A}_1,\dots,\hat{A}_n$
\cite{povm2},
\begin{equation}
\hat{\rho}(A_1,\dots,A_n)=\hat{K}(A_n)\hat{\rho}(A_1,\dots,A_{n-1})
\hat{K}^\dag(A_n)\;.
\end{equation}
The corresponding probability density is given by
$\rho(A_1,\dots,A_n)=\mathrm{Tr}\;\hat{\rho}(A_1,\dots,A_n)$.
We now define a family of Kraus operators,
namely $\hat{K}_\lambda(A) =
(2\lambda/\pi)^{1/4}\exp(-\lambda(\hat{A}-A)^2)$.  It is clear that
$\lambda\to\infty$ should correspond to exact, strong, projective
measurement, while $\lambda\to 0$ is a weak measurement and gives a
large error. We also see that strong projection changes the state (by
collapse), while $\lambda\to 0$ gives $\hat{\rho}(A)\sim \hat{\rho}$,
and hence this case corresponds to the weak measurement.  However, the
repetition of the same measurement $k$ times effectively means one
measurement with $\lambda\to k\lambda$ so, with $k\to\infty$, even a
weak coupling $\lambda\ll 1$ results in a strong measurement.  For an
arbitrary sequence of measurements, we can  write the final
density matrix as the convolution
\begin{equation}
  \hat{\rho}_{\boldsymbol\lambda}(\boldsymbol A) = 
  \int DA'\,\hat{\varrho}_{\boldsymbol\lambda}(\boldsymbol A')
  \prod_k g_k(A_k-A'_k)
\label{conv}
\end{equation}
with $g_k(A)= e^{-2\lambda_k A^2}\sqrt{2\lambda_k/\pi}$.  Here
$\boldsymbol \lambda=(\lambda_1,\dots,\lambda_n)$ $\boldsymbol
A=(A_1,\dots,A_n)$, and $DA=dA_1\dots dA_n$. The quasi-density matrix
$\hat{\varrho}$ is given recursively by
\begin{eqnarray}
  \hat{\varrho}_{\boldsymbol\lambda}(\boldsymbol A)  & = & 
  \int \frac{d\chi}{2\pi}e^{-i\chi A_n}
  \int\frac{d\phi}{\sqrt{2\pi\lambda_n}}e^{-\phi^2/2\lambda_n}\times
\label{quasi}\\
  &&
  e^{i(\chi/2+\phi)\hat{A}_n}\hat{\varrho}_{\boldsymbol\lambda}(A_1,\dots,A_{n-1})e^{i(\chi/2-\phi)\hat{A}_n}\nonumber
\end{eqnarray}
with the initial density matrix $\hat{\varrho}=\hat{\rho}$ for $n=0$.
We can interpret $g$ in (\ref{conv}) as some internal noise of the detectors
which, in the limit $\lambda\to 0$, should not influence the
system. We \emph{define} the quasiprobability $\varrho_{\boldsymbol
  \lambda}=\mathrm{Tr}\,\hat{\varrho}_{\boldsymbol \lambda}$ and
abbreviate $\varrho\equiv\varrho_{\boldsymbol 0}$. In this limit
(\ref{quasi}) reduces to
\begin{equation}
\hat{\varrho}(\boldsymbol A)=\int\frac{d\chi}{2\pi}e^{-i\chi A_n}
e^{i\chi\hat{A}_n/2}\hat{\varrho}(A_1,\dots,A_{n-1})e^{i\chi\hat{A}_n/2}\,.
\label{gene}
\end{equation}
Note that $\varrho_{0\dots 0,\lambda}=\varrho$, so the last
measurement does not need to be weak (it can be even a projection),
and marginal distributions are consistent with absence of a
measurement, $\int dA_k\,\varrho(\boldsymbol
A)=\varrho(\dots,A_{k-1},A_{k+1},\dots)$.  In the case of commuting
operators the quasiprobability reduces to the usual probability
$\varrho=\rho_{\boldsymbol\infty}$. For $\hat{A}_1=\hat{x}$ and $\hat{A}_2=\hat{p}$
with $[\hat{x},\hat{p}]=i\hbar$ we obtain the Wigner function
 $\varrho(x,p)=\varrho(p,x)=W(x,p)$ \cite{wigner}. 
The definition preserves locality -- for ($\hat{\boldsymbol A},\hat{\rho}_a$) and
($\hat{\boldsymbol B},\hat{\rho}_b$) acting in two separate Hilbert spaces we have
$\hat{\rho}=\hat{\rho}_a\hat{\rho}_b\to\varrho(\boldsymbol A,\boldsymbol B)=
\varrho_a(\boldsymbol A)\varrho_b(\boldsymbol B)$.
The averages with
respect to $\varrho$ are easily calculated by means of the generating
function (\ref{gene}), e.g. $\langle
A\rangle_\varrho=\mathrm{Tr}\,\hat{A}\hat{\rho}$, $\langle
AB\rangle_\varrho=\mathrm{Tr}\,\{\hat{A},\hat{B}\}\hat{\rho}/2$,
$\langle
ABC\rangle_\varrho=\mathrm{Tr}\,\hat{C}\{\hat{B},\{\hat{A},\hat{\rho}\}\}/4$
for $\boldsymbol A=(A,B,C)$. This ordering of operators is called
time symmetric \cite{schleich,hof}.  To relate the quasiprobability to
weak values, we have to consider two
measurements: $\hat{A}$ and $\hat{B}$.  The real part of the
weak value is just the average $\mathrm{Re}\:{}_B\langle
A\rangle_\rho=\langle A\rangle_{\varrho|B}$ with respect to the
conditional quasiprobability
$\varrho(A|B)=\varrho(A,B)/\varrho(B)$. The
complex weak values require a different interpretation \cite{aav},
which can also be generalized to sequential measurement \cite{mjp}.

We shall apply the above scheme to the measurement of
current $I(t)$ through a mesoscopic junction in a stationary state.
For a moment, we forget about quantum mechanics and
recall basic properties of stochastic processes \cite{kampen},
applying them to $\delta I=I-\langle I\rangle$.  It is convenient to
define the noise (second cumulant),
$\tilde{S}(\alpha,\beta)$ $= 2\pi\delta(\alpha+\beta) S_\alpha= \langle
\delta I(\alpha)\delta I(\beta)\rangle$, and the fourth cumulant ${\tilde
  C}(\alpha,\beta,\gamma,\eta)=
2\pi\delta(\alpha+\beta+\gamma+\eta)C(\alpha,\beta,\gamma,\eta)$ with
\begin{eqnarray}
&&{\tilde C}(\alpha,\beta,\gamma,\eta)=\langle \delta I(\alpha)\delta I(\beta)I(\gamma)\delta I(\eta)\rangle\\
&&
-\tilde{S}(\alpha,\beta)\tilde{S}(\gamma,\eta)
-\tilde{S}(\alpha,\gamma)\tilde{S}(\beta,\eta)
-\tilde{S}(\alpha,\eta)\tilde{S}(\gamma,\beta).\nonumber
\end{eqnarray}
Here and throughout the text we use Latin arguments in time domain and Greek ones in frequency domain, related by
$a(\omega)=\int dt\;e^{i\omega t}a(t)$. Note, that the delta function
of the frequency sum has a cut-off of the order of the 
measuring time $t_0$ (larger than all relevant timescales of the system), which in some following expressions is a
simple prefactor and does not enter final conclusions.


Let us define the fluctuating noise spectral density
$X_\omega = \int_{\omega_-}^{\omega_+} \delta I(\alpha)\delta
  I(-\alpha)d\alpha$ with $\omega_\pm=\omega\pm\delta\omega/2$,
for which we obtain the average fluctuations
\begin{eqnarray}
&&\langle(\delta X_\omega)^2\rangle/t_0=
\int_{\omega_-}^{\omega_+}
C_{\alpha\beta}d\alpha
d\beta+2\pi\int_{\omega_-}^{\omega_+}S^2_\alpha d\alpha,\nonumber\\
&&\langle\delta X_\omega\delta X_{\omega'}\rangle/t_0=
\int_{\omega_-}^{\omega_+}
d\alpha\int_{\omega'_-}^{\omega'_+}
C_{\alpha\beta}
d\beta
\end{eqnarray}
where $\delta X=X-\langle X\rangle$ and
$C_{\alpha\beta}=C(\alpha,-\alpha,\beta,-\beta)$.
The intervals $[\omega_-,\omega_+]$ and
$[\omega'_-,\omega'_+]$ are nonoverlapping.  Considering classical correlators of $\delta X$ at
different frequencies we obtain the Cauchy-Bunyakovsky-Schwarz
inequality
\begin{equation}
B=\frac{\langle\delta X_{\omega}\delta X_{\omega'}\rangle^2}{\langle(\delta X_{\omega})^2\rangle\langle(\delta X_{\omega'})^2\rangle}\leq 1.\label{cbs}
\end{equation}
If we choose e.g. $0\leq\omega'_-<\omega'_+<\omega_-$, the correlators correspond to
a low and high frequency measurement.  Furthermore, assuming that $S$
and $C$ are constant within the bandwidth $\delta\omega,\delta\omega'$, 
the inequality (\ref{cbs}) takes the form
(\ref{eq:1}) mentioned in the introduction. It is interesting to note,
that for frequency-independent (classical) noise the inequality is
always satisfied. 

Turning to the quantum case we stress that continuous measurement
cannot be strong as we would end up with the quantum Zeno effect and
suppress the dynamics of the system completely \cite{zeno}. This follows from
Eq.~ (\ref{quasi}) for Heisenberg operators $\hat{A}_k=\hat{A}(t_k=k\Delta)$ and
a finite $t_n=n\Delta$. In the limit $\sum\lambda_k\to\infty$ and
$n\to\infty$, $\hat\varrho$ becomes diagonal in the eigenbasis of
$\hat A$ and freezes. Nonclassical behavior of quantum correlations in
the limit of weak measurement can also be shown using the Leggett-Garg
inequality \cite{legarg}, which involves time-resolved second order
correlations assuming that the observables are bounded.  Our
inequality is more general as we do not require bounded observables
but we need fourth order correlations in the frequency domain instead,
which is more suitable for electric current measurements.

We denote the measured current operator in the Heisenberg picture by
$\hat{I}(t)$. Strong, projective measurement can be performed
only if we are interested in the long-time limit. The finer the time
resolution we want the weaker the measurement must be as
$[\hat{I}(t),\hat{I}(t')]\neq 0$.  We define a generating functional
of the quasiprobability in the weak measurement limit by
\begin{eqnarray}
  \varrho[I] & = & \int D\chi e^{-i\int \chi(t)I(t)dt}\label{vrho} \\\nonumber 
  &&\times \mathrm{Tr}
  \left[\mathcal Te^{\int
      i\chi(t)\hat{I}(t)dt/2}\hat{\rho}\tilde{\mathcal T}e^{\int
      i\chi(t)\hat{I}(t)dt/2}
  \right]
\end{eqnarray}
with $\mathcal T$ ($\tilde{\mathcal T}$) denoting (anti)time
ordering. This represents a straightforward generalization of the
generating function obtained from the trace of (\ref{gene}), in which
the time-variable labels the subsequent measurements.  The averages of
current powers (noise and third cumulant) have been already calculated
\cite{blanter,salo,zaikin1,Bednorz:09,les08,les09} with respect to the quasiprobability
(without introducing this notion) and measured
\cite{glatt-gab}. In experiments, the large measured offset
noise plays the role of $g$ in (\ref{conv}) preventing paradoxical results
\cite{ours}.  We emphasize that in long-time averages the
quasiprobability becomes a conventional probability and reproduces the formula for
the usual full counting statistics \cite{fcs}, also confirmed
experimentally \cite{glatt-schoel,reulet-rez}.

We consider a quantum point contact
described by fermionic operators around the Fermi level
\cite{blanter}.  Each operator $\hat{\psi}_{A\bar n}$ with $\bar
n=(n,\sigma)$ is denoted by mode number $n\in\{1..N\}$ and spin
orientation $\sigma$ and $A=L,R$ for left and right going electrons,
respectively.  Each mode can have its own Fermi velocity $v_n$ and
transmission coefficient $T_n$ (reflection $R_n=1-T_n$).  We will
assume non-interacting electrons and energy- and spin-independent
transmission through the junction.  The Hamiltonian including a
voltage bias $V$ reads
\begin{eqnarray}
  &&\hat{H}=\sum_{\bar n}\int dx\left\{i\hbar v_n[\hat{\psi}^\dag_{L\bar
    n}(x) \partial_x\hat{\psi}_{L\bar n}(x)- L\leftrightarrow R]\right.\nonumber\\
  &&+q_n\delta(x)[\hat{\psi}^\dag_{L\bar n}(x)\hat{\psi}_{R\bar n}(-x)+
    \hat{\psi}^\dag_{R\bar n}(x)\hat{\psi}_{L\bar n}(-x)]\nonumber\\
    &&\left.-eV\theta(x)[\hat{\psi}^\dag_{L\bar n}(x)\hat{\psi}_{L\bar n}(x)+
    \hat{\psi}^\dag_{R\bar n}(x)\hat{\psi}_{R\bar n}(x)]
  \right\}\,.
\end{eqnarray}
The fermionic operators satisfy anticommutation relations
$\{\hat{\psi}_{a}(x),\hat{\psi}_{b}(x')\}=0$ and
$\{\hat{\psi}_{a}(x),\hat{\psi}^\dag_{b}(x')\}=\delta_{ab}\delta(x-x')$
for $a,b=L\bar n,R\bar m$. The transmission coefficients are
$T_n=\cos^2(q_n/\hbar v_n)$. We apply (\ref{vrho}) to the current operator 
$\hat{I}=\sum_{\bar n}ev_n\hat{\psi}^\dag_{L\bar n}(0_+)
\hat{\psi}_{L\bar n}(0_+)-L\leftrightarrow R$ and the density
matrix $\hat{\rho}\propto\exp(-\hat{H}/k_BT)$.

\begin{figure}
  \includegraphics[scale=.5,angle=270]{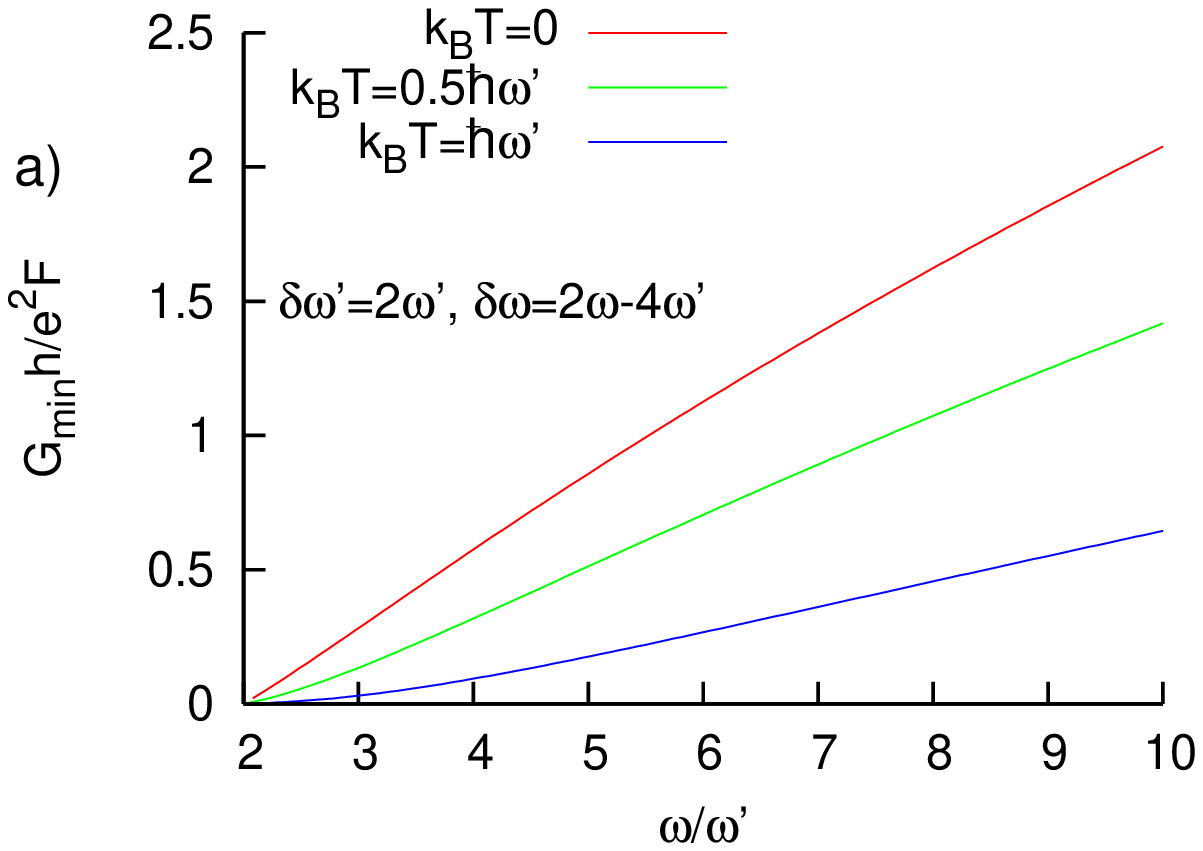}
\includegraphics[scale=.5,angle=270]{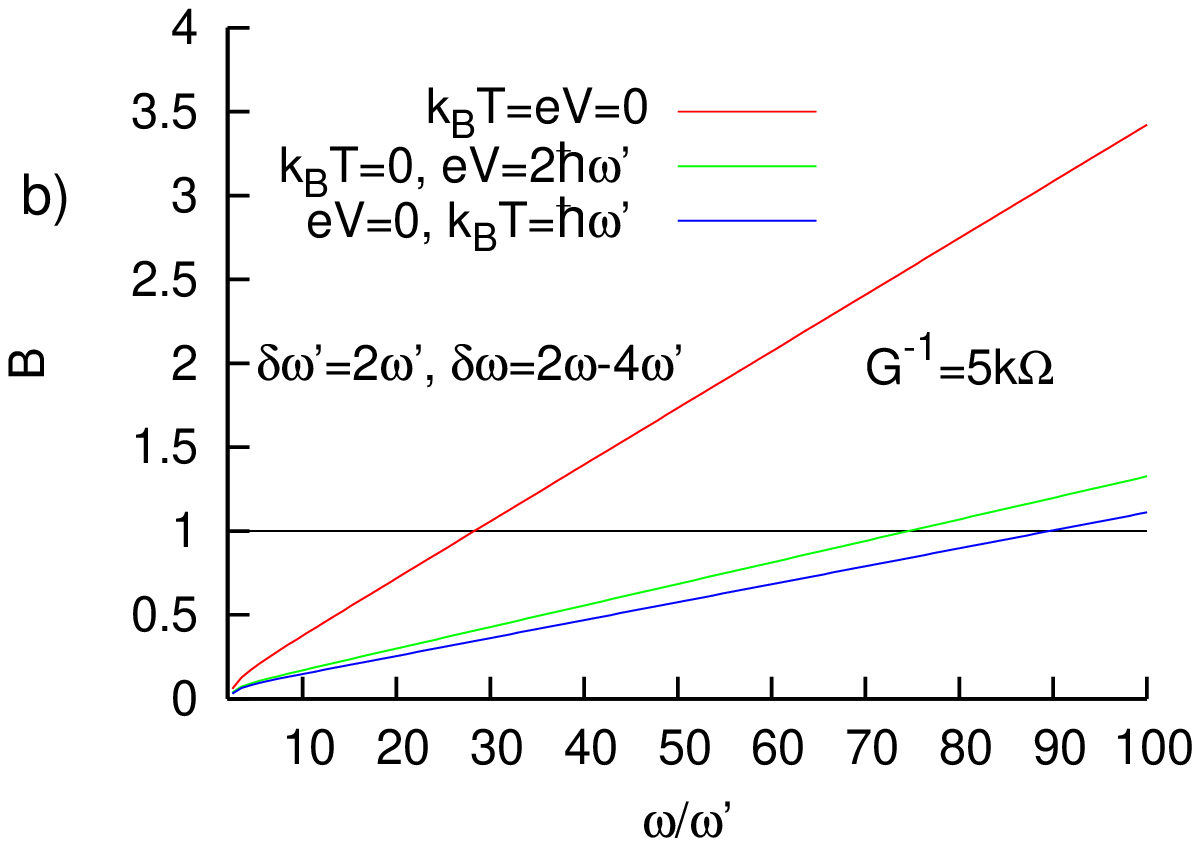}
  \caption{(color online) (a) The minimal value of the conductance $G$
    which satisfies the inequality (\ref{cbs}) for the mesoscopic
    junction at zero voltage.  The non-classicality occurs for
    $G<G_{min}$. (b)
 The dependence of inequality parameter $B$
    [given by (\ref{cbs})] for a the tunnel junction as function of
    the high frequency.  Below the line at $1$ is the classical
    regime. Either finite voltage and/or temperature lead to
    suppression of B and result in the necessity to measure at higher
    frequencies.}\label{gmin-bvd}
\end{figure}

To find conditions in which the inequality (\ref{eq:1}) is violated,
it is enough to consider the case $V=0$. Using Eq.~(\ref{vrho}) we obtain
 $S_\alpha=\hbar Gw(\alpha)$ and 
$ C_{\alpha\beta}=\hbar FGe^2 (w(\alpha)+w(\beta))/2$ \cite{blanter,zaikin1,Bednorz:09}. Here we denote
$w(\omega)=\omega\coth(\hbar\omega/2k_BT)$, conductance
$G=\sum_ne^2T_n/\pi\hbar$ and Fano factor
$F=\sum_nR_nT_n/\sum_nT_n$. In the tunneling limit, $T_n\ll 1$, for a
finite bias voltage $V$ we only need to replace $w(\omega)$ with
$(w(\omega+eV/\hbar)+w(\omega-eV/\hbar))/2$ and $F=1$.  For $T=35$mK,
$\delta\omega=\delta\omega'=2\omega'=2\pi\cdot 200$MHz,
$\omega=2\pi\cdot 6$GHz, $G^{-1}=500 \mathrm{k}\Omega$ and $V=0$, we
get $B=1.4$, which contradicts our classical expectation (\ref{cbs})
and clearly shows that the quasiprobability $\varrho$ must take
negative values. Generally, the violation occurs for sufficiently
small $G<G_{min}$, as shown in Fig.  \ref{gmin-bvd}a. At $eV=0$ and
$\delta\omega\simeq 2\omega\gg \delta \omega'\simeq 2\omega'\gtrsim
k_BT/\hbar$, we have $G_{min}=3\sqrt{\omega/\omega'}Fe^2/2h$.
For larger conductance, one can still find a reasonable range
of parameters for the violation at $G^{-1}=5\mathrm{k}\Omega$, as
shown in Fig. \ref{gmin-bvd}b. The strongest violation occurs at low
temperature and voltage but at large bandwidth. 
Unfortunately, the typical experimental conductance is with $G^{-1}\approx 50\Omega$
\cite{Reulet:10} even larger and would  require either $\omega\sim 2\pi\cdot
1$THz or a temperature $\sim 1$mK.  However, we can make the reasonable
assumption that all modes of the junction are independent and replace
the inequality (\ref{eq:1}) -- valid for the whole junction -- by the
same one for a single mode.  If we can assume that the modes are
independent and have similar transmission coefficient ($T_n\ll 1$) we can simply
divide $C$ and $S$ by the number of modes in (\ref{eq:1}). Note that
$C$ enters there linearly while $S$ enters quadratically, so
effectively we weaken the contribution from the second cumulant. For a
tunnel junction we can thus replace $G$ by $G/N_0$ where $N_0$ is a
lower bound of the contributing modes ($T_n>0$), which must be larger than $G\pi\hbar/e^2$.
The results in Fig.~\ref{gmin-bvd}b are valid for example for $N_0=100<
(h/2e^2)/50\Omega\approx 258$.  In this case, it is necessary to 
ensure that most of the modes contributing to the transport have small
transmission eigenvalues.

The cumulants are never measured directly.  The second cumulant always
contains a large offset noise generated by detector and amplifier and
effectively described by $\lambda$ in $g$. The offset noise can
presumably be subtracted as it is due to detector's amplifier. The
offset noise may be smaller in the case of a many-mode tunnel junction
and one way around is to measure cross correlations by different
detectors and amplifiers \cite{ours}.  The fourth cumulant should also
contribute to photon counting statistics \cite{schome,glatt3}, but in
the limit $\delta\omega\to 0$ the photon statistics is dominated by
the second cumulants in (\ref{eq:1}).

We have shown that the unusual properties of weak
measurements can be interpreted in terms of a real quasiprobability,
which can take negative values. Our interpretation agrees well with
predictions and measurements of the current fluctuations in mesoscopic
junctions. Its direct confirmation would be the measurement of
high-frequency fourth-order averages of the current though the
junction. By a violation of the inequality (\ref{eq:1}), the
negativity of the quasiprobability could be directly demonstrated.
Finally, the separation between detector and the system is somewhat
arbitrary.  One could argue that simply adding some noise can restore
the positive probability.  This is why an experimental estimate of
the detector noise $g$ or the strength of the measurement $\lambda$ is
also desirable.

We are grateful for fruitful discussions with and helpful suggestions
by J. Gabelli and B. Reulet.  Financial support from the DFG
through SFB 767 and SP 1285 is acknowledged.


\vspace{-5mm}
\begin{thebibliography}{99}
\bibitem{neumann}
J. von Neumann, \emph{Mathematical Foundations of Quantum Mechanics}
(Princeton U.P., Princeton, 1932).

\bibitem{kraus}
K. Kraus, \emph{States, Effects and Operations} (Springer, Berlin, 1983).
\bibitem{povm}
See e.g. J. Preskill, \textit{Quantum Information and Computation}
(www.theory.caltech.edu/people/preskill/ph229)
\bibitem{naim}
M. A. Naimark, Izv. Akad. Nauk. SSSR, Ser. Mat.
 {\bf 4}, 277 (1940).
\bibitem{brag} V. B. Braginsky and F. Ya. Khalili, \textit{Quantum Measurement} (Cambridge University Press, Cambridge, 1992).
\bibitem{jordan}
  A. N. Jordan and M. B\"uttiker, Phys. Rev. B {\bf 71}, 125333 (2005);
A. N. Jordan, A. N. Korotkov, and M. B\"uttiker, Phys. Rev. Lett. {\bf 97}, 026805 (2006);
A.N. Jordan and A.N. Korotkov, Phys. Rev. B {\bf 74}, 085307 (2006);
A. N. Korotkov in Quantum Noise in Mesoscopic Physics,
  Y.V. Nazarov (Ed.), (Kluwer, Dordrecht, 2003).
\bibitem{les08}
K. V. Bayandin, A.V. Lebedev, G. B. Lesovik, JETP {\bf 106}, 117 (2008).
\bibitem{aav}Y. Aharonov, D.Z. Albert, and L. Vaidman, Phys. Rev. Lett. {\bf 60}, 1351 (1988); Y. Aharonov and L. Vaidman, Phys. Rev. A {\bf 41}, 11 (1990);
Y. Aharonov and D. Rohrlich, \textit{Quantum Paradoxes} (Wiley-VCH, Weinheim, Germany, 2005).

\bibitem{wvopt}
G. J.~Pryde \textit{et al.},
Phys. Rev. Lett. {\bf 94}, 220405 (2005);
 A. M. Steinberg, Phys. Rev. Lett. {\bf 74}, 2405 (1995);
 H. M. Wiseman, Phys. Rev. A {\bf 65}, 032111 (2002).

\bibitem{wvcond}
 N. S.~Williams and A. N.~Jordan,
  Phys. Rev. Lett. {\bf 100}, 026804 (2008);
A. Romito, Y. Gefen, and Y.~M.~Blanter, 056801 (2008);
  V.~Shpitalnik, Y.~Gefen and A.~Romito, {\bf 101}, 226802 (2008).
\bibitem{glatt-gab}
E. Zakka-Bajjani \emph{et al.}, Phys. Rev. Lett. {\bf 99}, 236803 (2007);
{\bf 104}, 206802 (2010);
J. Gabelli and B. Reulet, {\bf 100}, 026601 (2008); J. Stat. Mech. P01049 (2009).
\bibitem{ours}
A. Bednorz and W. Belzig, Phys. Rev. Lett. {\bf 101}, 206803 (2008).
\bibitem{wigner}
E.P. Wigner, Phys. Rev. {\bf 40} 749 (1932); 
M. Hillery \textit{et al.},
Phys. Rep. {\bf 106}, 121 (1984).
\bibitem{povm2}
A. Schmid, Ann. of Phys. {\bf 173}. 103 (1987);
C. Anastopoulos and N. Savvidou, J. Math. Phys. {\bf 47}, 122106 (2006); {\bf 48}, 032106 (2007).
\bibitem{schleich}
  B. Berg \textit{et al.},
  Phys. Rev. A {\bf 80}, 033624 (2009).
\bibitem{hof}
H.F. Hofmann, Phys. Rev. A {\bf 81}, 012103 (2010).
\bibitem{mjp}
G. Mitchison, R. Jozsa, S. Popescu, Phys. Rev. A {\bf 76}, 062105 (2007).
\bibitem{kampen}
N.G. van Kampen, \textit{Stochastic Processes in Physics and Chemistry}
(North-Holland, Amsterdam, 2007).
\bibitem{zeno} B. Misra and E. C. G. Sudarshan, J. Math. Phys. (N.Y.) {\bf 18},
756 (1977).
\bibitem{legarg} A. J. Leggett and A. Garg, Phys. Rev. Lett. {\bf 54}, 857
(1985); A. J. Leggett, J. Phys. Condens. Matter {\bf 14}, R415
(2002).
 \bibitem{blanter}
Y.M. Blanter and M. B{\"u}ttiker, Phys. Rep. {\bf 336}, 1 (2000).
\bibitem{zaikin1}
A.V. Galaktionov, D.S. Golubev, and A.D. Zaikin, Phys. Rev. B. {\bf 68}, 235333 (2003);
{\bf 72}, 205417 (2005).
\bibitem{Bednorz:09} A. Bednorz and W. Belzig
Phys. Rev. B {\bf 81}, 125112 (2010).
\bibitem{salo}
J. Salo, F.~W.~J. Hekking, and J.~P. Pekola, Phys. Rev. B {\bf 74},
125427 (2006); T.~T. Heikkil\"a and T. Ojanen, Phys. Rev. B
\textbf{75}, 035335 (2007). 
\bibitem{les09}
S. Bachmann, G.~M. Graf, and G.~B. Lesovik, J. Stat. Phys {\bf 138}, 333 (2010).

\bibitem{fcs}
L.S. Levitov and G.B. Lesovik, JETP Lett. {\bf 58}, 230 (1993);
L.S. Levitov, H.W. Lee, G.B. Lesovik, J. Math. Phys. {\bf 37}, 4345 (1996);
W. Belzig and Y.V. Nazarov, Phys. Rev. Lett. {\bf 87}, 197006 (2001);
Y.V. Nazarov and M. Kindermann, Eur. J. Phys. B {\bf 35}, 413 (2003);
M. Kindermann and Y.V. Nazarov in Quantum Noise in Mesoscopic Physics,
  Y.V. Nazarov (Ed.), (Kluwer, Dordrecht, 2003).
\bibitem{glatt-schoel}
M. I. Reznikov \emph{et al.}, Phys. Rev. Lett. {\bf 75}  3340 (1995);
A. Kumar \emph{et al.}, {\bf 76}, 2778 (1996);
R.J. Schoelkopf \emph{et al.}, {\bf 78}, 3370 (1997).
\bibitem{reulet-rez}
B. Reulet \emph{et al.}, Phys. Rev. Lett. {\bf 91}, 196601 (2003);
Y. Bomze \emph{et al.}, {\bf 95}, 176601 (2005); {\bf 101}, 016803 (2008).
\bibitem{Reulet:10}
B. Reulet \emph{et al.}, in: \emph{Perspectives of Mesoscopic Physics},
A. Aharony, O. Entin-Wohlman (Eds.),
(World Scientific, Singapore, 2010)
\bibitem{schome}
C.W.J. Beenakker and H. Schomerus, Phys.Rev.Lett. {\bf 86}, 700 (2001);
{\bf 93}, 096801 (2004); A.V. Lebedev, G.B. Lesovik, and G. Blatter, Phys. Rev. B {\bf 81},
 155421 (2010).
\bibitem{glatt3}
J. Gabelli \emph{et al.}, Phys. Rev. Lett. {\bf 93} 056801 (2004).
\end{thebibliography}
\end{document}